\documentclass[3p, times,procedia]{elsarticle}

\usepackage{ecrc}   

\usepackage{ulem}

\volume{00}

\firstpage{1}

\runauth{}

\usepackage{amssymb}

\usepackage[figuresright]{rotating}

\begin{document}

\begin{frontmatter}

\dochead{}

\title{Precision X-ray spectroscopy of kaonic atoms as a probe of low-energy kaon-nucleus interaction}

\author[label1]{H.~Shi}
\ead{hexishi@lnf.infn.it} 
\author[label1]{M.~Bazzi}
\author[label2]{G.~Beer}
\author[label3,label4]{G.~Bellotti}
\author[label5,label1]{C.~Berucci}
\author[label1,label6]{A.~M.~Bragadireanu}
\author[label7]{D.~Bosnar}
\author[label5]{M.~Cargnelli}
\author[label1]{C.~Curceanu}
\author[label3,label4]{A.~D.~Butt}
\author[label1]{A.~d'Uffizi}
\author[label3,label4]{C.~Fiorini}
\author[label8]{F.~Ghio}
\author[label1]{C.~Guaraldo}
\author[label9]{R.S.~Hayano}
\author[label1]{M.~Iliescu}
\author[label5]{T.~Ishiwatari}
\author[label10]{M.~Iwasaki} 
\author[label1]{P.~Levi Sandri}
\author[label5]{J.~Marton}
\author[label10]{S.~Okada}
\author[label1,label6]{D.~Pietreanu}
\author[label1,label11]{K.~Piscicchia}
\author[label12]{A.~Romero Vidal}
\author[label1]{E.~Sbardella}
\author[label1]{A.~Scordo}
\author[label1,label6]{D.L.~Sirghi}
\author[label1,label6]{F.~Sirghi}
\author[label13,label14]{H.~Tatsuno}
\author[label15]{O.~Vazquez Doce}
\author[label5]{E.~Widmann} 
\author[label5]{J.~Zmeskal}

\address[label1]{INFN, Laboratori Nazionali di Frascati, Via E. Fermi 40, I-00044 Frascati(Roma), Italy,                               } 
\address[label2]{Department of Physics and Astronomy, University of Victoria, P.O. Box 1700 STN CNC, Victoria BC V8W 2Y2, Canada,      } 
\address[label3]{Politecnico di Milano, Dipartimento di Elettronica, Informazione e Bioingegneria, Milano, Italy,                      } 
\address[label4]{INFN Sezione di Milano, Via Celoria 16 - 20133 Milano, Italy,                                                         } 
\address[label5]{Stefan-Meyer-Institut f\"{u}r Subatomare Physik, Boltzmanngasse 3, 1090 Wien, Austria,                                } 
\address[label6]{IFIN-HH, Institutul National pentru Fizica si Inginerie Nucleara Horia Hulubbei, Reactorului 30, Magurele, Romania,   } 
\address[label7]{Department of Physics, Facaulty of Science, University of Zagreb, Bijenicka 32, HR-10000 Zagreb, Croatia,             } 
\address[label8]{INFN Sezione di Roma I and Instituto Superiore di Sanita, I-00161 Roma, Italy,                                        } 
\address[label9]{Department of Physics, School of Science, The University of Tokyo, Bunkyo-ku Hongo 7-3-1, Tokyo, Japan,               } 
\address[label10]{RIKEN, Institute of Physical and Chemical Research, 2-1 Hirosawa, Wako, Saitama 251-0198, Japan,                     }  
\address[label11]{Museo Storico della  Fisica e Centro Studi e Ricerche ``Enrico Fermi'', Piazza del Viminale 1-00184Roma, Italy,      }  
\address[label12]{Universidade de Santiago de Compostela, Casas Reais 8, 15782 Santiago de Compostela, Spain,                          }  
\address[label13]{National Institute of Standards and Technology (NIST), Boulder, CO, 80303, USA,                                      }  
\address[label14]{High Energy Accelerator Research Organization (KEK), Tsukuba, 305-0801, Japan,                                       }  
\address[label15]{Excellence Cluster Universe, Technische Universit\"{a}t M\"{u}nchen, Boltzmannstra\ss e 2, D-85748 Garching, Germany.}  


\begin{abstract}
In the exotic atoms where one atomic $1s$ electron is replaced by a $K^{-}$, 
the strong interaction between the $K^{-}$ and the nucleus introduces an energy shift and broadening of the low-lying kaonic atomic levels 
which are determined by only the electromagnetic interaction. 
By performing X-ray spectroscopy for Z=1,2 kaonic atoms, 
the SIDDHARTA experiment determined with high precision the shift and width for the $1s$ state of $K^{-}p$ and the $2p$ state of kaonic helium-3 and kaonic helium-4. 
These results provided unique information of the kaon-nucleus interaction in the low energy limit. 
}
\end{abstract}


\end{frontmatter}

\graphicspath{{Figures/}}

\section{Introduction}
\label{intro}
Precision X-ray spectroscopy of the Z=1,2 kaonic atoms holds the key to understand the low-energy interaction between the kaon and the nucleus. 
In kaonic hydrogen which is the simplest kaonic atom,
the strong interaction induced shift and width of the $1s$ state can be deduced from the $K$-series X-rays,
and they are related to the $s$-wave $K^{-}p$ scattering length $a_{K^{-}p}$ through the Deser-Trueman formula \cite{Des54}. 
This scattering length consists of two isospin dependent components of $a_{0}$ (I=0) and $a_{1}$ (I=1), 
which are among the most fundamental parameters for the low-energy $\bar{K}N$ interaction, and which can only be derived experimentally from X-ray spectroscopy. 
To disentangle these two components,
precise measurement of the kaonic deuterium $1s$ shift and width is necessary, 
further taking into account higher order contributions from the three-body interaction \cite{Mei04}.
Experimentally this measurement is extremely difficult mainly due to the small X-ray yield and the broad width of the $1s$ level. 
Until now an unambiguous detection of the kaonic deuterium X-rays has not been achieved and the measurement remains to be the major challenge of upcoming experiments.

For the Z=2 kaonic atoms, there had been a long-standing discrepancy between the theoretical and the measured strong-interaction induced shift of the $2p$ level of kaonic helium-4. 
This "kaonic helium puzzle" was solved by the KEK E570 experiment which showed the $2p$ level shift was 2 $\pm$ 2(stat) $\pm$ 2(syst) eV \cite{Oka07}, 
refuting the previous average shift of about - 40 eV from three early experiments \cite{Wie71, Bat79, Bai83}. 
More essentially, 
the precise measurement of the kaonic helium $2p$ level shift is a test ground for possible deeply-bound kaonic nuclear states predicted by Akaishi and Yamazaki \cite{Aka02, Aka05}. 
A series of electronvolt precision measurement of the shift and the width of the $2p$ level for both kaonic helium-3 and kaonic helium-4 atoms 
can effectively narrow down the allowed range of the nuclear potential of kaon in the search of deeply-bound states.

The measurements we have performed in the SIDDHARTA experiment will be introduced in detail next.

\section{The SIDDHARTA experiment}

\begin{figure}[htbp]
\centering
\includegraphics[width=8cm,clip]{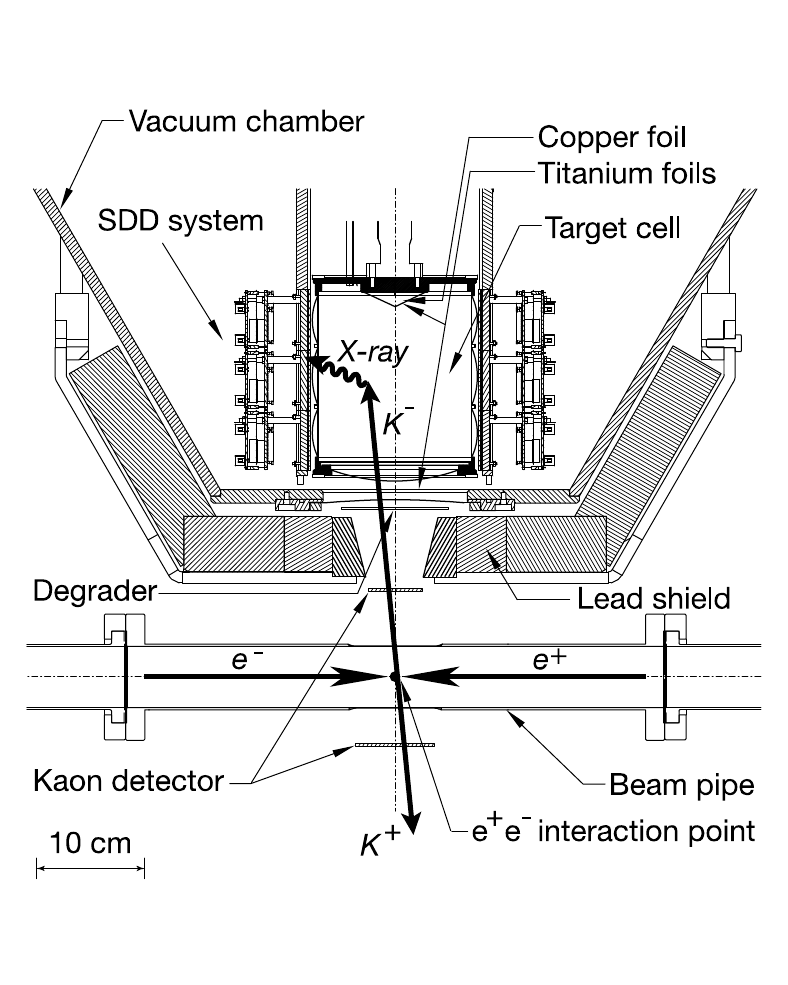}
\caption{A schematic cutaway view of the SIDDHARTA setup at the DA$\Phi$NE interaction point. 
The charged kaon pairs are identified with two plastic scintillators, 
and the $K^{-}$ induced X-rays detected by the SDDs are identified from the time correlation to the kaon pair events. 
}
\label{fig-1}       
\end{figure}

\subsection{Experimental method}
The experiment was performed at the DA$\Phi$NE $e^{-}e^{+}$ collider in Frascati, Italy. 
The beam energies are tuned to produce $\phi$ (1020) resonance at rest in the lab frame, 
and then with a 49\% branching ratio the $\phi$ decays into a pair of $K^{+}K^{-}$. 
As the kaon detectors, we used two plastic scintillators placed above and below the interaction point as illustrated in Fig. \ref{fig-1}. 
The kaons are distinguished from the minimum ionizing particles using the time of flight information at the kaon detectors.
And then from the coincidence of the two scinillators we defined the kaon trigger, which marks the timing of the incident kaons.

A fraction of the negatively charged kaons that made the kaon trigger will stop inside the volume of the gaseous target 
placed about 20 cm above the interaction point to form kaonic atoms.
With the time correlation between the subsequent X-rays detected by the SDDs and the kaon trigger,
one can distinguish the kaon origin X-rays from the background coming from the processes that are asynchronous to the kaon pair production. 

With a total active area of 144 cm$^2$, the SDDs cover about 10\% of the solid angle around the target cell. 
The energy calibration of the SDDs were done every few hours by shining Ti and Cu foils with an X-ray tube. 
Using the $K_{\alpha}$ lines of Ti (4.5 keV) and Cu (8.0 keV), we determined the scale of the energy spectra, 
and the energy resolution at 6 keV was stable at about 150 eV (FWHM) throughout our measurement. 
At the operating temperature of 140 K, the SDDs showed a timing resolution of 700 FWHM, 
effectively rejected the asynchronous background by four orders of magnitude. 
The configuration and the performance of the detectors are discussed in detail in previous publications of the SIDDHARTA experiment \cite{Baz09, BazKp}.

During the data taking in 2009, we accumulated data with gaseous targets of hydrogen (1.3 g/l), deuterium (2.50 g/l), 
helium-3 (0.96 g/l), and helium-4 (1.65 g/l and 2.15 g/l).

\subsection{Kaonic hydrogen and deuterium}
The main objective of the SIDDHARTA experiment is the precise measurement of the kaonic hydrogen $K$-series X-rays, 
thus most of the beam time was dedicated to the hydrogen target measurement to a total intergrated luminosity of 340 pb$^{-1}$. 
A total amount of 100 pb$^{-1}$ integrated luminosity was dedicated to the first exploratory kaonic deuterium measurement. 

In the final analysis to derive the $1s$ shift $\epsilon_{1s}$ and width $\Gamma_{1s}$ of kaonic hydrogen, 
the kaonic X-ray spectra from the two data sets were fitted simultaneously.
As the yield of the kaonic deuterium $K$-series X-rays is more than one order of magnitude smaller than that of the kaonic hydrogen, 
the spectrum from the deuterium target measurement helped determining the background X-rays from other kaonic atoms when the $K^{-}$ stopped inside the target cell made of Kapton. 
This combined analysis contributed to reduce the systematic errors in the kaonic hydrogen results, which achieved the best precision up to date for the $1s$ shift  
$\epsilon_{1s}$ = -283 $\pm$ 36 (stat.) $\pm$ 6 (syst.) eV and for the width $\Gamma_{1s}$ = 541 $\pm$ 89 (stat.) $\pm$ 22 (syst.) eV \cite{BazKp}.
The results solved the discrepancy between two recent kaonic hydrogen X-ray experiments $KpX$ in KEK \cite{KpX} and DEAR at DA$\Phi$NE \cite{DEAR}, 
in that the two results do not overlap with each other within error bars. 
The improved precision presents more stringent constrains to the theoretical study of the low-energy QCD near the $K^{-}p$ threshold. 

From the deuterium target spectrum which showed no significant amount of signal of kaonic deuterium X-rays, 
we evaluated an upper limit for the yield of the kaonic deuterium $K_{\alpha}$ X-ray as $Y(K_{\alpha})$ < 0.0039 
(C.L. 90\%) \cite{BazKd}.
It presents an important reference to plan for the future precision measurement of kaonic deuterium X-rays.

\subsection{Kaonic helium-3 and helium-4}
The strong interaction induced $2p$ level shift and width are measured for both kaonic helium-3 and helium-4 atoms, 
and it is the first time that kaonic helium-3 X-rays are measured. 
The results of the shift as listed in \ref{tab-1} are consistent with the results of E570 \cite{Oka07}, 
thus a zero-compatible shift of the $2p$ level from experiment is established, which is in agreement with the theoretical estimations in \cite{Bai83, Bat90}. 
We have not found abnormally large widths which 
can directly support 
the estimations made in conjunction with the prediction of a deeply-bound kaon states\cite{Aka02, Aka05}. 

\begin{table}[h]
\centering
\caption{Results on the energy shifts ($\Delta E_{2p}$) and widths ($\Gamma_{2p}$) of the kaonic helium-3 and kaonic helium-4 $2p$ states \cite{Baz09, Baz12}.}
\label{tab-1}      
\begin{tabular}{ccc}
\hline
\hline
Target & $\Delta E_{2p}$ [{\rm eV}] & $\Gamma_{2p}$ [eV] \\\hline
helium-4 & +5 $\pm$ 3 (stat.) $\pm$ 4 (syst.) & 14 $\pm$ 8 (stat.) $\pm$ 5 (syst.) \\
helium-3 & -2 $\pm$ 2 (stat.) $\pm$ 4 (syst.) & \quad 6 $\pm$ 6 (stat.) $\pm$ 7 (syst.) \\\hline
\end{tabular}
\end{table}


\section{Future perspectives}
Following the method successfully developed in the SIDDHARTA experiment, 
the extended SIDDHARTA-2 collaboration is in preparation of a series of upgrades and modifications of the apparaturs, 
aiming at the first precision measurement of kaonic deuterium X-rays \cite{Sidt2}.

\paragraph{Acknowledgements}
We thank C. Capoccia, G. Corradi, B. Dulach, and D. Tagnani from LNF-INFN; 
and H. Schneider, L. Stohwasser, and D. St\"{u}kler from Stefan-Meyer-Institut, 
for their fundamental contribution in designing and building the SIDDHARTA setup. 
We thank as well the DA$\Phi$NE staff for the excellent working conditions and permanent support. 
Part of this work was supported by the European Community-Research Infrastructure 
Integrating Activity ``Study of Strongly Interacting Matter" (HadronPhysics2, 
Grant Agreement No. 227431, and HadronPhysics3 (HP3) Contract No. 283286) 
under the Seventh Framework Programme of EU; 
HadronPhysics I3 FP6 European Community program, Contract No. RII3-CT-2004- 506078; 
Austrian Science Fund (FWF) (P24756-N20); 
Austrian Federal Ministry of Science and Research BMBWK 650962/0001 VI/2/2009; 
Croatian Science Foundation under Project No. 1680;
Romanian National Authority for Scientific Research, 
Contract No. 2-CeX 06-11-11/2006; 
and the Grant-in-Aid for Specially Promoted Research (20002003), MEXT, Japan.

\bibliographystyle{elsarticle-num}

\end{document}